\documentclass[10pt]{article}                           

\def \Oeuvres{O$\!$euvres}
\def \ie {i.e.~} 
\def\LHS{l.h.s.~}
\def \ccomma{\raise 2pt\hbox{,}} 
\def \D {\hbox{d}}
\def \Log {\mathop{\rm Log}\nolimits}
\def \Transp#1{{}^{\rm t} #1}
\def\CRAS{C.~R.~Acad.~Sc.~Paris}
\def\SAM{Stud.~Appl.~Math.~}
\def\Pn     {{\rm Pn}}
\def\PI     {{\rm P1}}
\def\PII    {{\rm P2}}
\def\PIII   {{\rm P3}}
\def\PIV    {{\rm P4}}
\def\PV     {{\rm P5}}
\def\PVI    {{\rm P6}}

\def\Alpha{A}
\def\Beta {B}
\def\abcd{\alpha,\beta,\gamma,\delta}
\def\ABCD{\Alpha,\Beta,\Gamma,\Delta}

\def\veca{\mbox{\boldmath{$\alpha$}}}
\def\vecA{\mbox{\boldmath{$\Alpha$}}}
\def\vect{\mbox{\boldmath{$\theta$}}}
\def\vecT{\mbox{\boldmath{$\Theta$}}}

\def\BiT{birational transformation}

\def \TG      {{\rm T}_{\rm G}}     
\def \TPVI    {{\rm T}_{\rm 6}}     %
\def \TPVIb   {{\rm T}_{\rm 6,b}}   
\def \TPVb    {{\rm T}_{\rm 5,b}}   
\def \TPIVb   {{\rm T}_{\rm 4,b}}   
\def \TPVIu   {{\rm T}_{\rm 6,u}}   
\def \TPVu    {{\rm T}_{\rm 5,u}}   
\def \TPIVu   {{\rm T}_{\rm 4,u}}   

\def\Sa  {{\rm S}_{\rm a}}
\def\Sb  {{\rm S}_{\rm b}}
\def\Sc  {{\rm S}_{\rm c}}
\def\Sd  {{\rm S}_{\rm d}}

\def \Hbadc {{\rm H}_{\rm badc}}  
\def \Hdcba {{\rm H}_{\rm dcba}}  
\def \Hcdab {{\rm H}_{\rm cdab}}  

\def\Mugan{Mu\u gan}

\def \H {{\rm H}}  

\def\vplus {     \bar{v}}
\def\vminus{\underbar{$v$}}

\begin{document}

\title{
 First degree birational transformations of the Painlev\'e equations
and their contiguity relations
\footnote{S2001/017. To appear in J.~Phys.~A, special issue SIDE IV, Tokyo,
27 November-1 December 2000.
\hfill \break
Corresponding author RC, fax +33--1--69088786, phone +33--1--69087349.
nlin.SI/0110028
}
}

\author{Robert Conte\dag\ and Micheline Musette\ddag
{}\\
\\ \dag Service de physique de l'\'etat condens\'e, CEA--Saclay
\\ F--91191 Gif-sur-Yvette Cedex, France
\\ E-mail:  Conte@drecam.saclay.cea.fr
{}\\
\\ \ddag Dienst Theoretische Natuurkunde, Vrije Universiteit Brussel
\\ Pleinlaan 2, B--1050 Brussels, Belgium
\\ E-mail:  MMusette@vub.ac.be 
}

\maketitle

\hfill 

{\vglue -10.0 truemm}
{\vskip -10.0 truemm}

\begin{abstract} 
 We present a consistent truncation,
allowing us to obtain the first degree birational transformation 
found by Okamoto for the sixth Painlev\'e equation.
The discrete equation arising from its contiguity relation is then
just the sum of six simple poles.
An algebraic solution is presented,
which is equivalent to but simpler than the Umemura solution.
Finally, the well known confluence provides a unified picture of all 
first degree birational transformations for the lower Painlev\'e equations,
ranging them in two distinct sequences.

\end{abstract}


\noindent \textit{Keywords}:
Painlev\'e equations,
birational transformation,
contiguity relation,
singular manifold method,
truncation.

\noindent \textit{PACS}
 02.30.+g   

\baselineskip=12truept 


\tableofcontents

\vfill \eject

\section{Introduction}

The purpose of this article is to derive 
\textit{birational transformations} for the Painlev\'e equations $\Pn$
by a direct method, only based on the singularity structure of the 
ordinary differential equation (ODE).
By definition,
a birational transformation is a set of two relations,
\begin{eqnarray} 
& &
u = f(U',U,X),\
U = F(u',u,x),
\label{eqbira}
\end{eqnarray}
with $f$ and $F$ rational functions,
which maps an equation
\begin{eqnarray} 
& &
E(u) \equiv \Pn(u,x,\veca)=0,\
\veca=(\abcd),\
\label{eqPnu}
\end{eqnarray}
into the same equation with different parameters
\begin{eqnarray} 
& &
E(U) \equiv \Pn(U,X,\vecA)=0,\
\vecA=(\ABCD),
\label{eqPnU}
\end{eqnarray}
with some homography (usually the identity) between $x$ and $X$.
The parameters $(\veca,\vecA)$ must obey as many algebraic relations 
as elements in $\veca$.
The \textit{degree} of a \BiT\ is defined as the highest degree in $U'$
or $u'$ of the numerator and the denominator of (\ref{eqbira}).

Since the sixth of these Painlev\'e equations $\Pn$ generates
the five others by a confluence process \cite{PaiCRAS1906},
we first concentrate on it
\begin{eqnarray*}
\label{eqP6}
\PVI\ : \
u''
&=&
\frac{1}{2} \left[\frac{1}{u} + \frac{1}{u-1} + \frac{1}{u-x} \right] {u'}^2
- \left[\frac{1}{x} + \frac{1}{x-1} + \frac{1}{u-x} \right] u'
\\
& &
+ \frac{u (u-1) (u-x)}{x^2 (x-1)^2} 
  \left[\alpha + \beta \frac{x}{u^2} + \gamma \frac{x-1}{(u-1)^2} 
        + \delta \frac{x (x-1)}{(u-x)^2} \right]\cdot
\end{eqnarray*}

Schlesinger \cite[p.~144]{SchlesingerP6}
was the first to prove the existence of a transformation breaking the
invariance under permutation of the four singular points $\infty,0,1,x$.
To express it,
one needs to introduce the \textit{monodromy exponents} 
(according to the classical terminology \cite{FuchsP6}),
\ie\ the four components of the vectors
\begin{eqnarray}
& &
\vect=\pmatrix{\theta_\infty \cr \theta_0 \cr \theta_1 \cr \theta_x},\
\vecT=\pmatrix{\Theta_\infty \cr \Theta_0 \cr \Theta_1 \cr \Theta_x},\
\end{eqnarray}
only defined by their squares,
\begin{eqnarray} 
& &
\theta_\infty^2= 2 \alpha,\
\theta_0^2     =-2 \beta,\
\theta_1^2     = 2 \gamma,\
\theta_x^2     =1 - 2 \delta,
\\
& &
\Theta_\infty^2= 2 \Alpha,\
\Theta_0^2     =-2 \Beta,\
\Theta_1^2     = 2 \Gamma,\
\Theta_x^2     =1 - 2 \Delta.
\end{eqnarray}
The transformation found by Schlesinger
conserves two monodromy exponents and shifts the two others by one unit,
up to sign changes of course,
\begin{eqnarray} 
& &
{\hskip -10.0truemm}
\theta_i=\Theta_i,\
\theta_j=\Theta_j,\
\theta_k=\Theta_k+1,\
\theta_l=\Theta_l+1,\
(i,j,k,l) \hbox{ permutation of }
(\infty,0,1,x).
\label{eqSchlesingerShifts}
\end{eqnarray}
However, Schlesinger did not give the associated birational representation.

This was achieved by Garnier \cite{Garnier1943a,Garnier1951},
to establish the missing proof of a theorem of Schwarz on the problem of 
Plateau.
The \BiT\ which realizes the shifts (\ref{eqSchlesingerShifts})
is of second degree,
\begin{eqnarray}
& &
\frac{u U}{x} = \frac{R_n^{+} R_n^{-}}{R_d^{+} R_d^{-}}\ccomma\
x=X,\
\label{eqTGu}
\\
& &
R_n^{\pm} =
\frac{x (x-1) U'}{U(U-1)(U-x)}
 + \frac{\pm \Theta_0}{U}
 + \frac{\Theta_1}{U-1}
 + \frac{\Theta_x -1}{U-x}\ccomma
\\
& &
R_d^{\pm} =
\frac{x (x-1) U'}{U(U-1)(U-x)}
 + \frac{\pm \Theta_\infty-\Theta_1-\Theta_x+1}{U}
 + \frac{\Theta_1}{U-1}
 + \frac{\Theta_x -1}{U-x}\ccomma
\\
& &
\TG :
 \vect = \pmatrix{\Theta_\infty \cr \Theta_0 \cr -\Theta_1 \cr -\Theta_x}
       + \pmatrix{0 \cr 0 \cr 1 \cr 1 \cr}.
\label{eqTG}
\end{eqnarray}
Since the transformation between $\vect$ and $\vecT$ is an involution,
the inverse of (\ref{eqTGu}) is readily obtained by exchanging
$(u,\vect)$ and $(U,\vecT)$.
We will adopt such a convention (choice of signs so as to have involutions)
throughout the present paper.

This birational transformation was later rediscovered by several authors 
\cite{FY,Okamoto1987I,MS1995b}.
Garnier added that the relation (\ref{eqTGu})
can be considered as a generalization of the 
\textit{contiguity relation} for $\PVI$,
thus extending the well known similar notion introduced by Gauss for the
hypergeometric equation.
However, (\ref{eqTGu}) is not a ``pure'' contiguity relation,
but a two-point relation also involving derivatives.
A three-point relation without derivatives can be found in Section
\ref{sectionDiscrete}.

Another, elegant way of finding {\BiT}s has been devised by 
Okamoto \cite{Okamoto1987I}.
The Hamiltonian of $\PVI$ found by Malmquist \cite{MalmquistP6}
has no such invariance as (\ref{eqSchlesingerShifts}),
but a quite simple (in fact, affine) transformation
makes it invariant under any permutation of four components $b_j$ 
different from (but equivalent to) $\theta_j$, defined as
\begin{eqnarray}
& &
b_1=(\theta_0    + \theta_1)/2,\
b_2=(\theta_0    - \theta_1)/2,\
b_3=(\theta_x -1 - \theta_\infty)/2,\
b_4=(\theta_x -1 + \theta_\infty)/2.
\label{eqBasisbOkamoto}
\label{eqBasisb}
\end{eqnarray}
This allowed Okamoto to build various {\BiT}s,
by construction canonical in the Hamiltonian sense.
One of them (his ``parallel transformation'')
is identical to that of Garnier.
Another one
(denoted $w_1 w_2 w_1$ in Ref.~\cite[p.~356 ex.~2.1]{Okamoto1987I}
 and $\TPVI$ here)
seems to have degree two but in fact has degree one.
Indeed, the first line of the matrix equation which defines it is
\begin{eqnarray}
& &
q_w = q + \frac{(b_3-b_1)q (q-1) p}{\D h / \D t+b_1^2}\ccomma
\end{eqnarray}
with the correspondence of notation $q_w=u,q=U$ and $p$ linear in our $U'$,
and, with the expression for $\D h / \D t$ on last line of page 356
[in which a factor $q$ has been omitted, see
 the $\PVI$ Hamiltonian page 339, repeated page 348],
a common factor $p$ to the numerator and the denominator cancels out,
\begin{eqnarray}
& &
q_w = q + \frac{(b_3-b_1)q (q-1)}{-q(q-1) p +[b_1(2q-1)-b_2]}\ccomma
\end{eqnarray}
yielding a first degree transformation.
This transformation
is more elementary in the sense that
the transformation of Garnier is an integer power \cite{C2001TSP6}
of $\TPVI$.
This transformation $\TPVI$ has recently been rediscovered \cite{NJH},
up to a simultaneous homography on $(u,x)$.

To achieve our goal (rely only on the singularity structure to find {\BiT}s),
we need to improve the \textit{singular manifold method}
so that it succeeds to obtain a \BiT\ for $\PVI$ and, by confluence,
for each $\Pn$ equation, $n=5,4,3,2$
($\PI$ depends on no parameter and thus admits no \BiT).
Originally introduced for partial differential equations (PDEs)
by Weiss, Tabor and Carnevale \cite{WTC},
the singular manifold method is a powerful tool
for deriving B\"acklund transformations,
by considering only the singularity structure of the solutions.
Its current achievements are detailed in summer school proceedings,
see Refs.~\cite{Cargese1996Musette,CetraroConte}.
An extension to ODEs has been proposed \cite{CJP,GJP1999a} 
to derive a \BiT\ for the Painlev\'e equations,
but its application to the master equation $\PVI$ is still an open problem.
We solve it here by implementing an essential piece of information,
which has up to now been overlooked.

The paper is organized as follows.
In Section \ref{sectionHomography},
we exploit the information that there always exists a homography between 
the derivative 
of the solution 
of the considered Painlev\'e equation $\Pn$
and the Riccati pseudopotential $Z$
introduced in the ``truncation'' assumption.
This reduces the problem to finding two functions of two variables
instead of two functions of three variables.

In Section \ref{sectionOneFamilyTruncation},
we implement this homography in the one-family truncation,
which allows us to overcome the major difficulty, coming, 
in the case of $\PVI$, from the value $1$ of the Fuchs index.
We also improve a previous conjecture \cite{FA1982}
on the necessary form of the \BiT.
The truncation then becomes easy to solve and,
up to the four homographies on $(U,x)$ which conserve $x$,
it admits a solution, 
identical to the transformation $w_1 w_2 w_1$ of Okamoto.

In Section \ref{sectionAfirst degreeBiTofP6},
we give the various representations of this transformation.

In Section \ref{sectionFixedPoints},
by looking for the fixed points of the {\BiT}s of $\PVI$,
we obtain a quite simple algebraic solution.

In Section \ref{sectionConfluence},
starting from the \BiT\ of $\PVI$,
we perform the classical confluence,
and obtain two sequences of first degree {\BiT}s for the lower $\Pn$ 
equations.

In Section \ref{sectionDiscrete},
we solve the recurrence relation between the monodromy exponents of $\PVI$,
and we build the contiguity relation,
depending on four arbitrary parameters.


In a previously submitted article
\cite{CM2001alreadysubmitted},
we have already presented several results,
in particular the detailed algorithm of the truncation
(which we do not repeat here)
and its result for $\PVI$.
What is new here is the first coherent description of all the first degree
birational transformations of all the $\Pn$ equations,
obtained by the simultaneous application of two noncommuting operations,
the confluence from $\Pn$ to a lower $\Pn$
and the homographies conserving $\Pn$.
This unified picture in turn generates a similar unification of all the 
contiguity relations (discrete equations)
which arise from a birational transformation.


Throughout the paper,
we evidently consider as identical two {\BiT}s which only differ 
by sign changes and homographies.
Moreover, without loss of generality,
our {\BiT}s are always involutions for the same, consistent, choice of signs.

\section{The fundamental homography}
\label{sectionHomography}

{}From the result of Richard Fuchs \cite{FuchsP6},
the $\PVI$ equation is obtained from the zero curvature condition
of a linear differential system of order two.
Therefore the pseudopotential of the singular manifold method
has only one component \cite{MC1991}, 
which can be chosen so as to satisfy some Riccati ODE.

Each $\Pn$ equation which admits a \BiT\ has one or several
(four for $\PVI$)
couples of families of movable simple poles with opposite residues $\pm u_0$,
therefore both the one-family truncation and the two-family truncation
are applicable.
In the present paper,
we consider only the one-family truncation,
whose assumption is \cite{GJP1999a}
\begin{eqnarray} 
& &
u=u_0 Z^{-1} +U,\ u_0 \not=0,\
x=X,
\label{eqDTOne}
\\
& &
Z'=1 + z_1 Z + z_2 Z^2,\ z_2 \not=0,
\label{eqRiccatiOneFamily}
\end{eqnarray}
in which $u$ and $U$ satisfy (\ref{eqPnu}) and (\ref{eqPnU}),
$(Z,z_1,z_2)$ 
are rational functions of $(x,U,U')$ to be determined.
After this is done, the relation (\ref{eqDTOne}) 
represents half of the birational transformation.

Besides the equation (\ref{eqRiccatiOneFamily}),
there exists a second Riccati equation in the present problem,
this is the Painlev\'e equation (\ref{eqPnU}) itself.
Indeed, any $N$-th order, first degree ODE with the Painlev\'e property
is necessarily \cite[pp.~396--409]{PaiLecons}
a Riccati equation for $U^{(N-1)}$,
with coefficients depending on $x$ and the lower derivatives of $U$,
in our case
\begin{eqnarray}
& &
U''=A_2(U,x) U'^2 + A_1(U,x) U' + A_0(U,x).
\label{eqRiccatiUprime}
\end{eqnarray}

Since the group of invariance of a Riccati equation is the homographic group,
the variables $U'$ and $Z$ are linked by a homography,
the three coefficients $g_j$ of which are rational in $(U,x)$.
Let us define it as
\begin{eqnarray}
& &
 (U' + g_2) (Z^{-1} - g_1) - g_0=0,\ g_0 \not=0.
\label{eqHomographyRUprime}
\end{eqnarray}

This allows us to obtain the two coefficients
 $z_j$ of the Riccati pseudopotential equation (\ref{eqRiccatiOneFamily})
as explicit expressions 
of $(g_j, \partial_U g_j, \partial_x g_j, A_2,A_1,A_0,U')$.
Indeed,
eliminating $U'$ between (\ref{eqRiccatiUprime}) and
(\ref{eqHomographyRUprime}) defines a first order ODE for $Z$,
whose identification with (\ref{eqRiccatiOneFamily})
\textit{modulo} (\ref{eqHomographyRUprime})
provides three relations.

For the one-family truncation, these are
\begin{eqnarray} 
& &
g_0= g_2^2 A_2 - g_2 A_1 + A_0 + \partial_x g_2 - g_2 \partial_U g_2,
\label{eqg0Apriori}
\\
& &
z_1=
   A_1 - 2 g_1 + \partial_U g_2 - \partial_x \Log g_0 
+ \left(2 A_2 - \partial_U \Log g_0\right) U',
\label{eqz1Apriori}
\\
& &
z_2= -g_1 z_1 - g_1^2 - g_0 A_2 - \partial_x g_1 - (\partial_U g_1) U'.
\label{eqz2Apriori}
\end{eqnarray}

Therefore, the natural unknowns in the present problem are the two
coefficients $g_1,g_2$ of the homography,
which are functions of the two variables $(U,x)$,
and not the two functions $(z_1,z_2)$ of the three variables $(U',U,x)$.

\textit{Remark}.
One must also consider the case when the relation between $Z^{-1}$ and $U'$
is affine,
excluded in (\ref{eqHomographyRUprime}).
Assuming
\begin{eqnarray}
& &
G_1 (U' + G_2) - Z^{-1} =0,\ G_1 \not=0,
\label{eqLinearZUprime}
\end{eqnarray}
the equation analogous to (\ref{eqg0Apriori}) is now
\begin{eqnarray}
& &
\partial_U G_1 + G_1^2 + A_2 G_1=0,
\end{eqnarray}
which for $\PVI$ admits no solution $G_1$ rational in $U$.

\section{The truncation}                                  
\label{sectionOneFamilyTruncation}

Just like the field $u$ is represented, see Eq.~(\ref{eqDTOne}),
by a Laurent series in $Z$ which terminates (``truncated series''),
the \LHS\ $E(u)$ of the $\Pn$ equation 
can also be written as a truncated series in $Z$.
This is achieved 
by the elimination of $u,Z',U'',U'$
between (\ref{eqPnu}), (\ref{eqPnU}), (\ref{eqDTOne}), 
(\ref{eqRiccatiOneFamily})
and (\ref{eqHomographyRUprime}),
followed by the elimination of $(g_0,z_1,z_2)$
from (\ref{eqg0Apriori})--(\ref{eqz2Apriori})
($q$ denotes the singularity order of $\Pn$ written as a 
differential polynomial in $u$, it is $-6$ for $\PVI$),
\begin{eqnarray}
& &
E(u) = \sum_{j=0}^{-q+2} E_j(U,x,u_0,g_1,g_2,\veca,\vecA) Z^{j+q-2}=0,
\label{eqST1LaurentE}
\\
& &
\forall j\ :\            E_j(U,x,u_0,g_1,g_2,\veca,\vecA)=0.
\label{eqST1Determining}
\end{eqnarray}
The nonlinear \textit{determining equations} $E_j=0$ are independent of $U'$,
and this is the main difference with previous work \cite{GJP1999a}.
Another difference is the greater number ($-q+3$ instead of $-q+1$)
of equations $E_j=0$,
which is due to the additional elimination of $U'$ with 
(\ref{eqHomographyRUprime}).

The $-q+3$ determining equations (\ref{eqST1Determining}) 
in the three unknown functions $u_0(x)$, $g_1(U,x)$, $g_2(U,x)$ 
(and the unknown scalars $\abcd$ in terms of $\ABCD$)
must be solved,
as usual, by increasing values of their index $j$.

For any solution $g_j(U,x)$, there exist evidently three other solutions,
generated by the action on $g_j(U,x)$ of the 
three homographies of $U$ which conserve $x$ and $\PVI$, namely
\begin{eqnarray}
& &
x=X,\
\left\lbrace
\matrix{
\displaystyle{
\Hdcba\ : \ 
u-x=\frac{x(x-1)}{U-x}\ccomma
}
\hfill \cr
\displaystyle{
\Hbadc\ : \
u=\frac{x}{U}\ccomma
}
\hfill \cr
\displaystyle{
\Hcdab\ : \
u-1=\frac{1-x}{U-1}\cdot
}
\hfill \cr}
\right.
\end{eqnarray}

In order to shorten the resolution,
it is advisable to enforce the condition of birationality,
which requires that $g_k,k=0,1,2,$ be the quotient of two polynomials of $U$.
After their degree has been found,
the equations (\ref{eqST1Determining}) split into an even more overdetermined
set of equations involving functions of $x$ only, much easier to solve.

But the decisive shortening results from the following stronger condition.
{}From the expression of the direct \BiT,
\begin{eqnarray}
& &
u=U + u_0 \left(g_1(U,x) + \frac{g_0(U,x)}{U'+g_2(U,x)}\right),
\label{eqSTDirecte}
\end{eqnarray}
the movable values of $U$ which make the field $u$ infinite 
are defined (apart from the poles arising from the terms $U$, $g_1$, and $g_0$,
which are fixed)
by the ODE
\begin{eqnarray}
& &
U'+g_2(U,x)=0.
\label{eqRiccatiDenom}
\end{eqnarray}
For all known {\BiT}s of $\Pn$
(see the book \cite{GLBook} for $\PII$ to $\PV$,
      Ref.~\cite[formula (2.8)]{Garnier1951}
  and Ref.~\cite[p.~356]{Okamoto1987I} for $\PVI$),
it happens that the ODE analogous to (\ref{eqRiccatiDenom}) is a Riccati ODE
(or a product of Riccati ODEs in \cite{Garnier1951}),
\ie\ the unique first order first degree ODE which has the 
Painlev\'e property.
In at least one other example of higher order \cite{Hone1998HH},
the birational transformation between two different 
ODEs having the Painlev\'e property
has a denominator which defines a $\PI$ equation.
Let us conjecture the generality of this property.
\medskip

\noindent \textbf{Conjecture}. 
\textit{Given a birational transformation between two ODEs having
the Painlev\'e property,
the ODEs defined by its denominators also have the Painlev\'e property.
}
\medskip

This is an improvement of a previous conjecture by Fokas and Ablowitz
\cite[formula (2.6)]{FA1982}
in two respects:
both fields $u$ and $U$ are required to satisfy the same ODE,
and no specific $U$-dependence is assumed for $g_0$ and $g_1$;
their conjecture happens to be true for $\PII$-$\PV$ but not for $\PVI$.

The practical resolution for $\PVI$ is performed in 
\cite{CM2001alreadysubmitted}
and, \textit{modulo} the homographies on $U$ which conserve $x$,
one finds the solution
\begin{eqnarray}
{\hskip -7.0 truemm}
g_0
& = &
\frac{N U (U-1)(U-x)}{u_0 x(x-1)},\
g_1=0,\
g_2=\frac{U (U-1)(U-x)}{x (x-1)}
\left(\frac{\Theta_0}{U}+\frac{\Theta_1}{U-1}+\frac{\Theta_x-1}{U-x}\right)
\ccomma
\label{eqoldeq27}
\\
{\hskip -7.0 truemm}
u_0
& = &
-\frac{x(x-1)}{\theta_\infty}\ccomma\
\theta_\infty = \frac{1}{2}
  \left( \Theta_\infty - \Theta_0 - \Theta_1 - \Theta_x + 1\right),
\label{eqg0g1g2}
\\
{\hskip -7.0 truemm}
N
& = &
1-\Theta_\infty-\Theta_0-\Theta_1-\Theta_x,
\label{eqold29}
\\
{\hskip -7.0 truemm}
z_1 
& = &
 \frac{1}{x(x-1)} ((\Theta_1  +\Theta_x-1)  U
                  +(\Theta_x-1+\Theta_0  ) (U-1)
                  +(\Theta_0  +\Theta_1  ) (U-x))
+\frac{1}{x}+\frac{1}{x-1}\ccomma
\label{eqz1sol}
\\
{\hskip -7.0 truemm}
z_2
& = &
   \frac{N \theta_\infty}{2(x(x-1))^2} ((U-1)(U-x)+U(U-x)+U(U-1)).
\label{eqz2sol}
\end{eqnarray}

Although, by lack of time, we have not yet examined all the subcases of this
resolution,
this solution is quite probably unique.

\section{A first degree \BiT\ of $\PVI$}
\label{sectionAfirst degreeBiTofP6}

Let us denote it $\TPVI$.
The eight signs $s_\infty,s_0,s_1,s_x$ and $S_\infty,S_0,S_1,S_x$,
with $s_j^2=S_j^2=1$,
of the monodromy exponents remain arbitrary and independent.
Whenever there is no ambiguity, $\TPVI$ will also denote the case with all
$+1$ signs.

The \textit{affine representation} of $\TPVI$ is
\begin{eqnarray}
& &
\TPVI : 
s_j \theta_j = 
S_j \Theta_j - \frac{1}{2} \left(\sum S_k \Theta_k\right) + \frac{1}{2}
\ccomma
\label{eqT6}
\\
& &
\TPVI^{-1}:
S_j \Theta_j = 
s_j \theta_j - \frac{1}{2} \left(\sum s_k \theta_k\right) + \frac{1}{2}
\ccomma
\label{eqT9}
\end{eqnarray}
in which $j,k=\infty,0,1,x$.
The convention adopted for the signs is aimed at making $\TPVI$
equal to its inverse when the signs verify $S_j=s_j$.

Denoting $N$ the odd-parity constant
\begin{eqnarray}
N
& = &
\sum (\theta_k^2 - \Theta_k^2)
\label{eqP6N}
\\
& = &
  1 - \sum S_k \Theta_k
=
 -1 + \sum s_k \theta_k
\label{eqNvect}
\\
& = &
  2 (s_j \theta_j - S_j \Theta_j),\ j=\infty,0,1,x,
\label{eqNvecT}
\end{eqnarray}
the \textit{birational representation} is (for clarity, the signs are omitted)
\begin{eqnarray}
\frac{N}{u-U}
& = &
  \frac{x (x-1) U'}{U (U-1)(U-x)}
 +\frac{\Theta_0}{U}+\frac{\Theta_1}{U-1}+\frac{\Theta_x-1}{U-x}
\label{eqTP6uvecTUnsigned}
\\
& = &
  \frac{x(x-1)u'}{u(u-1)(u-x)}
 +\frac{\theta_0}{u}+\frac{\theta_1}{u-1}+\frac{\theta_x-1}{u-x}\cdot
\label{eqTP6uvectUnsigned}
\end{eqnarray}

The four algebraic relations between $\abcd$ and $\ABCD$,
equivalent to (\ref{eqT6}), are
\begin{eqnarray}
& &
\forall j=\infty,0,1,x\ : \
(\theta_j^2+ \Theta_j^2 - (N/2)^2)^2 - (2 \theta_j \Theta_j)^2=0.
\end{eqnarray}

\section{Fixed points of the \BiT\ of $\PVI$}
\label{sectionFixedPoints}

Every fixed point of every \BiT\ generically defines 
a zero-parameter algebraic solution in the following way.
Given a \BiT\ ${\rm T}$,
its fixed points $\vect$ are by definition all the solutions of the affine
matrix equation
\begin{eqnarray}
& &
{\rm T}
\pmatrix{\theta_\infty \cr
         \theta_0      \cr
         \theta_1      \cr
         \theta_x      \cr}
=
{\rm P}
\pmatrix{s_\infty \theta_\infty \cr
         s_0      \theta_0      \cr
         s_1      \theta_1      \cr
         s_x      \theta_x      \cr},
\end{eqnarray}
in the unknowns 
${\rm P}$ (a permutation matrix),
$s_i$ (four independent signs),
$\theta_i$ (four complex numbers).
The solution $u$ is then the common solution to $\PVI$ and to 
(\ref{eqbira}) with $u=U$,
and therefore it is generically a zero-parameter algebraic solution of $\PVI$.

Let us review these fixed points for the three classes of {\BiT}s:
the three homographies on $(u,x)$ (apart the identity) which conserve $x$,
the \BiT\ of Garnier,
the first degree \BiT.

For the three homographies on $(u,x)$ which conserve $x$,
one obtains the zero-parameter algebraic solutions
\begin{eqnarray}
& &
u=
\left\lbrace
\matrix{
\displaystyle{
x + \sqrt{x(x-1)},\
}
& \vect^2=\Transp (\lambda^2,\mu^2,\mu^2,\lambda^2),\
\hfill \cr
\displaystyle{
\sqrt{x},\
}
& \vect^2=\Transp (\lambda^2,\lambda^2,\mu^2,\mu^2),\
\hfill \cr
\displaystyle{
1 + \sqrt{1-x},\
}
& \vect^2=\Transp (\lambda^2,\mu^2,\lambda^2,\mu^2).
\hfill \cr}
\right.
\label{eqP6solnew}
\end{eqnarray}
It is convenient to introduce the crossratio $(x_1,x_2,x_3,x_4)$
of four numbers,
\begin{eqnarray}
& &
(x_1,x_2,x_3,x_4)
=\frac{(x_3-x_1)(x_4-x_2)}{(x_3-x_2)(x_4-x_1)}\ccomma
\label{eqcrossratio}
\end{eqnarray}
and to denote $a_i,a_j,a_k,a_l$ an arbitrary permutation of 
the four singular points $\infty,0,1,x$ of $\PVI$.
Note that $(\infty,0,1,z)=z$.
The unique expression of the above zero-parameter algebraic solution
is then
\begin{eqnarray}
& &
(a_i,a_j,a_k,u)=\sqrt{(a_i,a_j,a_k,a_l)},\
\theta_i^2=\theta_j^2=\lambda^2,\
\theta_k^2=\theta_l^2=\mu^2.
\label{eqP6solalgebraic}
\end{eqnarray}
Curiously, we could not find this quite elementary solution in 
the classical authors (Painlev\'e, Boutroux, Garnier, Malmquist).
The first mention of its existence seems to be in \cite[Eq.~(4.4)]{Maz2001b}.

For the second degree \BiT\ of Garnier,
there exists only one fixed point,
found by Umemura \cite{Umemura1998},
which satisfies (\ref{eqTG}) with $\vecT=\vect$.
This solution
(two components of $\vect$ equal to $\pm 1/2$, the two others arbitrary)
corresponds to two zero-parameter algebraic solutions,
either (as originally written by Umemura)
\begin{eqnarray}
& &
u=
\left\lbrace
\matrix{
\displaystyle{
 \frac{\lambda^2 x + \lambda \mu \sqrt{x(x-1)}}
      {\lambda^2 x + \mu^2 (1-x)}\ccomma
}
& 
\vect^2=\Transp (1/4,\lambda^2,\mu^2,1/4),\
\hfill \cr
\displaystyle{
 \frac{(\lambda^2-\mu^2) x + \lambda \mu (x-1) \sqrt{x}}
      {\lambda^2 x - \mu^2}\ccomma
}
&
\vect^2=\Transp (1/4,1/4,\mu^2,\lambda^2),\
\hfill \cr
\displaystyle{
 \frac{\lambda^2 x - \lambda \mu x \sqrt{1-x}}
      {\lambda^2 - \mu^2 (1-x)}\ccomma
}
&
\vect^2=\Transp (1/4,\lambda^2,1/4,\mu^2),\
\hfill \cr}
\right.
\label{eqP6solUmemurawithden}
\end{eqnarray}
or
\begin{eqnarray}
& &
u=
\left\lbrace
\matrix{
\displaystyle{
 x - \frac{\lambda}{\mu} \sqrt{x(x-1)},
}
& 
\vect^2=\Transp (\lambda^2,1/4,1/4,\mu^2),\
\hfill \cr
\displaystyle{
 -\frac{\lambda}{\mu} \sqrt{x},
}
&
\vect^2=\Transp (\lambda^2,\mu^2,1/4,1/4),\
\hfill \cr
\displaystyle{
 1 + \frac{\mu}{\lambda} \sqrt{1-x},
}
&
\vect^2=\Transp (\lambda^2,1/4,\mu^2,1/4),\
\hfill \cr}
\right.
\label{eqP6solUmemurawithoutden}
\end{eqnarray}
in which $(\lambda,\mu)$ are arbitrary.

For the first degree \BiT\ ${\rm T}=\TPVI$,
the property that $\sum \theta_j^2$ is conserved is equivalent to the property
that $\sum s_j \theta_j$ is unity.
Therefore, from Eq.~(\ref{eqNvecT}),
there is a unique fixed point,
characterized by the simultaneous vanishing of both sides of
(\ref{eqTP6uvecTUnsigned}),
this is the set of one-parameter solutions of the Riccati equation,
first found by R.~Fuchs.

Under the action of $\TPVI$ (\textit{modulo} signs and homographies),
the algebraic solution (\ref{eqP6solnew}) is mapped 
either to (\ref{eqP6solUmemurawithden})
or     to (\ref{eqP6solUmemurawithoutden}).
These two solutions, equivalent under $\TPVI$,
are inequivalent under the \BiT\ of Garnier.

Since the set (\ref{eqP6solnew}) 
and the sets (\ref{eqP6solUmemurawithden}), (\ref{eqP6solUmemurawithoutden})
belong to the same equivalence class,
it is advisable to choose (\ref{eqP6solnew}) as the representative of the
solution of Umemura.

\section{Confluence to first degree birational transformations}
\label{sectionConfluence}

Let us show that \textit{all} first degree {\BiT}s
of the lower $\Pn$ equations can be generated from the first degree \BiT\
$\TPVI$ of $\PVI$.

Let us first define the $\Pn$ equations which admit a \BiT\ as
\begin{eqnarray*}
\PVI\ : \
u''
&=&
\frac{1}{2} \left[\frac{1}{u} + \frac{1}{u-1} + \frac{1}{u-x} \right] {u'}^2
- \left[\frac{1}{x} + \frac{1}{x-1} + \frac{1}{u-x} \right] u'
\\
& &
+ \frac{u (u-1) (u-x)}{x^2 (x-1)^2} 
  \left[\alpha + \beta \frac{x}{u^2} + \gamma \frac{x-1}{(u-1)^2} 
        + \delta \frac{x (x-1)}{(u-x)^2} \right]\ccomma
\\
\PV\ : \
u''
&=&
\left[\frac{1}{2 u} + \frac{1}{u-1} \right] {u'}^2
- \frac{u'}{x}
+ \frac{(u-1)^2}{x^2} \left[ \alpha u + \frac{\beta}{u} \right]
+ \gamma \frac{u}{x}
+ \delta \frac{u(u+1)}{u-1}\ccomma
\nonumber
\\
\PIV\ : \
u''
&=&
\frac{u'^2}{2 u} + \gamma \left(\frac{3}{2} u^3 + 4 x u^2 + 2 x^2 u\right)
- 2 \alpha u + \frac{\beta}{u}\ccomma
\\
\PIII'\ : \
u''
&=&
\frac{u'^2}{u} - \frac{u'}{x} + \frac{\alpha u^2 + \gamma u^3}{4 x^2}
 + \frac{\beta}{4 x}
 + \frac{\delta}{4 u}\ccomma
\\
\PII\ : \
u''
&=&
\delta (2 u^3 + x u)
 + \alpha.
\end{eqnarray*}
As compared to the usual choice \cite{GambierThese}
\begin{eqnarray*}
   & & \PIV\ : \ \gamma=1,
\\ & & \PIII (u,x     ,\alpha,\beta,\gamma,\delta)
      =\PIII'(x u, x^2,\alpha,\beta,\gamma,\delta),
\\ & & \PII\ : \ \delta=1,
\end{eqnarray*}
the additional symbols $\gamma$ in $\PIV$ and $\delta$ in $\PII$
have been added to represent the signs of the two opposite residues $\pm u_0$
of each family of movable simple poles.
The $\PIII'$ equation is that defined by Painlev\'e
\cite[p.~1115]{PaiCRAS1906} in the class of $\PIII$.
Table \ref{Table0} collects the relevant data,
in particular the definition of the monodromy exponents.

\textit{Remark}.
Those of the parameters $\veca$ which are unused in the definition of 
the monodromy exponents $\vect$ in Table \ref{Table0} 
are necessarily invariant under any \BiT.
Thus,
$\Delta=\delta$ for $n=5,3,2$,
$\Gamma=\gamma$ for $n=  4,3$.


\begin{table}[h] 
\caption[garbage]{
Definition of the monodromy exponents and other useful data 
for the $\Pn$ equations.
We follow the notation of Okamoto \cite{Okamoto1986Pn},
rather than the one of Jimbo and Miwa \cite{JimboMiwaII},
and each monodromy exponent $\theta_j$, including $\theta_\infty$,
has a square rational in $\abcd$.
The successive lines are~:
the singularity order $q$ of the $\Pn$ equation and the positive 
Fuchs index $i$,
the value of the first coefficient $u_0$ of the Laurent series for $u$,
the notation for the square root of $u_0$,
the definition of the monodromy exponents $\vect$,
the components of the column vector $\vect$.
{}
}
\vspace{0.2truecm}
\begin{center}
\begin{tabular}{| l | l | l | l | l | l |}
\hline 
&
$\PII$   
&
$\PIII'$ 
&
$\PIV$   
&
$\PV$    
&
$\PVI$   
\\ \hline 
$q,i$
&
$-3,4$          
&
$-4,2$          
&
$-4,3$          
&
$-5,1$          
&
$-6,1$          
\\ \hline 
$u_0$
&
$d^{-1}$           
&
$2 c^{-1} x$       
&
$c^{-1}$           
&
$\theta_\infty^{-1} x$ 
&
$\theta_\infty^{-1} x (1-x)$   
\\ \hline 
$\sqrt{\phantom{x}}$
&
$d^2=\delta$          
&
$c^2=\gamma$          
&
$c^2=\gamma$          
&
$\theta_\infty^2=2 \alpha$        
&
$\theta_\infty^2=2 \alpha$        
\\ \hline 
$\alpha$
&
$ - d \theta_\infty$     
&
$ 2 c \theta_\infty$     
&
$ 2 c \theta_\infty$     
&
$     \theta_\infty^2/2$ 
&
$     \theta_\infty^2/2$ 
\\ 
$\beta$
&
&
$ - 2 d \theta_0$        
&
$ - 8   \theta_0^2$      
&
$ -     \theta_0^2/2$    
&
$ -     \theta_0^2/2$    
\\ 
$\gamma$
&
&
$c^2$                     
&
$c^2$                     
&
$ - d \theta_1 $          
&
$     \theta_1^2/2$       
\\ 
$\delta$
&
$d^2$                     
&
$-d^2$                    
&
&
$-d^2/2$                  
&
$(1-\theta_x^2)/2$        
\\ \hline 
$\vect$
&
$\pmatrix {\theta_\infty \cr}$                             
&
$\pmatrix {\theta_\infty \cr \theta_{0} \cr}$              
&
$\pmatrix {\theta_\infty \cr \theta_{0} \cr}$              
&
$\pmatrix {\theta_\infty \cr \theta_{0} \cr \theta_1 \cr}$ 
&
$\pmatrix {\theta_\infty \cr \theta_{0} \cr \theta_{1} \cr \theta_{x} \cr}$
\\ \hline 
\end{tabular}
\end{center}
\label{Table0}
\end{table}

The successive coalescences of the four singular points of $\PVI$
\cite{PaiCRAS1906}
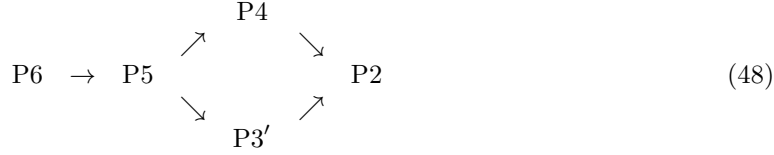
\begin{equation} 
\matrix{
     &             &     &          & \PIV   &          &      & \cr
     &             &     & \nearrow &        & \searrow &      & \cr
\PVI & \rightarrow & \PV &          &        &          & \PII & \cr
     &             &     & \searrow &        & \nearrow &      & \cr
     &             &     &          & \PIII' &          &      & \cr
} 
\end{equation}
from an equation $E(x,u,\abcd)=0$ 
  to an equation $E(X,U,\ABCD)=0$
are described by Poincar\'e perturbations 
$(x,u,\abcd,\vect) \to (X,U,\ABCD,\vecT,\varepsilon), \varepsilon \to 0$.
The confluence formulae for $\abcd$ are classical \cite{PaiCRAS1906},
those for the monodromy exponents $\vect$ have been established
in \cite{Okamoto1986Pn}, 
they are recalled in Table \ref{TableConfluence}.
All these transformations are affine,
and this is the reason why Painlev\'e introduced $\PIII'$ to replace 
his original choice of $\PIII$.

\tabcolsep=1.5truemm
\tabcolsep=0.5truemm

\begin{table}[h] 
\caption[garbage]{
Confluence of the monodromy exponents.
The parameters $c,d$ (which essentially represent signs) also participate to
the confluence. 
The choice of square roots is such that 
there are only $+$ signs in the successive values
$
 \theta_\infty+ \theta_0+\theta_1+\theta_x,\
 \theta_\infty+ \theta_0+\theta_1,\
2\theta_\infty+2\theta_0,\
 \theta_\infty+ \theta_0,\
2\theta_\infty$.
{}
}
\vspace{0.2truecm}
\begin{center}
\begin{tabular}{| l | l | l | l | l | l | l | l | l |}
\hline 
&
$x$
&
$u$
&
$\theta_\infty$
&
$\theta_0$
&
$\theta_1$
&
$\theta_x$
&
$c$
&
$d$
\\ \hline 
$6 \to 5$
&
$1+\varepsilon X$
&
$U$
&
$\theta_\infty$
&
$\theta_0$
&
$\theta_1 - \varepsilon^{-1} d$
&
$\varepsilon^{-1} d + \varepsilon$
&
&
\\ \hline 
$5 \to 4$
&
$1+ \varepsilon X$
&
$\varepsilon U/2$
&
$ -2 c \varepsilon^{-2}$
&
$ 2 \theta_0$
&
$ 2 \theta_\infty + 2 c \varepsilon^{-2}$
&
&
&
$ 2 c \varepsilon^{-2} - 2 \theta_\infty$ 
\\ \hline 
$5 \to 3'$
&
$X$
&
$1 + \varepsilon U$
&
$\theta_\infty + \varepsilon^{-1}c/2$
&
$-\varepsilon^{-1}c/2$
&
$\theta_0$
&
&
&
$ \varepsilon d/2$ 
\\ \hline 
$4 \to 2$
&
$ \varepsilon X/4 - \varepsilon^{-1}$
&
$\varepsilon^{-1} + U$
&
$ - \varepsilon^{-3} d$
&
$\theta_\infty + \varepsilon^{-3} d$
&
&
&
$4 \varepsilon^{-1} d$ 
&
\\ \hline 
$3' \to 2$
&
$1+\varepsilon^2 X/2$
&
$1 + \varepsilon U$
&
$ 4 d \varepsilon^{-3}$
&
$2 \theta_\infty -4 d \varepsilon^{-3}$
&
&
&
$-4 d \varepsilon^{-3}$
&
$ 2 \theta_\infty + 4 d \varepsilon^{-3}$ 
\\ \hline 
\end{tabular}
\end{center}
\label{TableConfluence}
\end{table}

Let us first remark that the confluence cannot increase the degree of the \BiT.

When acting on $\TPVI$, the confluence generates two inequivalent 
first degree {\BiT}s at the $\PV$ level.
Indeed, at the $\PVI$ level the four singular points are equivalent
but, for going to the $\PV$ level
(two equivalent singularities plus another one,
chosen once and for all as $(\infty,0)$ and $1$),
there exist two inequivalent choices of confluence,
and both must be performed.
The first choice, which we call ``normal'',
is the one adopted in Table \ref{TableConfluence},
which selects $(x,1)$ as the coalescing pair.
The second choice, called ``biased'',
is to select a pair made of one point among $(\infty,0)$
and a second point among $(1,x)$.
A given choice is equivalent to performing the corresponding homography
prior to the standard confluence.
Among the four homographies which conserve $x$ and $\PVI$,
two (the identity and $\Hbadc$) correspond to the first choice
and two ($\Hdcba$ and $\Hcdab$) to the second choice.
To maximize the symmetry,
let us choose the biased transformation $\TPVIb$
as the product of $\TPVI$ on the right by $\Hdcba$,
with signs reversals for the $\infty$ and $x$ components 
($\Sa,\Sb,\Sc,\Sd$ denote the operators which change the sign of, 
respectively, $\theta_\infty$, $\theta_0$, $\theta_1$, $\theta_x$),
\begin{eqnarray}
& &
\TPVIb=\Sa \Sd \TPVI \Hdcba \Sd \Sa.
\end{eqnarray}
Its affine and birational representations are
\begin{eqnarray}
& &
\PVI\ : \
\pmatrix{s_\infty \theta_\infty \cr
         s_0      \theta_0      \cr
         s_1      \theta_1      \cr
         s_x      \theta_x      \cr}
=-\frac{1}{2}              \pmatrix{ 1 & -1 & -1 & -1 \cr
                                    -1 &  1 & -1 & -1 \cr
                                    -1 & -1 &  1 & -1 \cr
                                    -1 & -1 & -1 &  1 \cr}
\pmatrix{S_\infty \Theta_\infty \cr
         S_0      \Theta_0      \cr
         S_1      \Theta_1      \cr
         S_x      \Theta_x      \cr}
              + \frac{1}{2} \pmatrix{-1 \cr 1 \cr 1 \cr -1 \cr}\ccomma
\label{eqT6biasedAffine0}
\end{eqnarray}
and
\begin{eqnarray}
\PVI
& : &
\frac{-N x (x-1)}{(u-x)(U-x)-x(x-1)}
= 
(U-x) \times
\nonumber
\\
& &
\phantom{xxxxxxxxxxxxxxx}
\left(  \frac{x (x-1) U'}{U (U-1)(U-x)}
 +\frac{\Theta_0}{U}+\frac{\Theta_1}{U-1}
 +\frac{\Theta_\infty-\Theta_0-\Theta_1}{U-x}
\right)
\ccomma
\\
& &
\frac{N x (x-1)}{(u-x)(U-x)-x(x-1)}
= 
(u-x) \times
\nonumber
\\
& &
\phantom{xxxxxxxxxxxxxxx}
\left(  \frac{x (x-1) u'}{u (u-1)(u-x)}
 +\frac{\theta_0}{u}+\frac{\theta_1}{u-1}
 +\frac{\theta_\infty-\theta_0-\theta_1}{u-x}
\right)
\ccomma
\end{eqnarray}
in which the odd-parity constant $N$,
again equal to the difference of the squared norms of the monodromy exponents,
as in Eq.~(\ref{eqP6N}),
now takes different affine expressions in $(\theta_j,\Theta_j)$,
\begin{eqnarray}
N
& = &
\sum (\theta_k^2 - \Theta_k^2)
\label{eqP6Nbiased0}
\\
& = &
  1 + S_\infty \Theta_\infty - S_0 \Theta_0 - S_1 \Theta_1 + S_x \Theta_x
\\
& = &
 -1 - s_\infty \theta_\infty + s_0 \theta_0 + s_1 \theta_1 - s_x \theta_x
\\
& = &
  -2 (s_\infty \theta_\infty - S_x \Theta_x)
=  2 (s_0      \theta_0      - S_1 \Theta_1)
=  2 (s_1      \theta_1      - S_0 \Theta_0)
= -2 (s_x      \theta_x      - S_\infty \Theta_\infty).
\label{eqP6Nbiased4}
\end{eqnarray}

Therefore, under the standard confluence,
$\TPVI$ and $\TPVIb$ generate two \textit{inequivalent}
first degree {\BiT}s at the $\PV$ level.
The sum $ \theta_\infty+ \theta_0+\theta_1+\theta_x$ goes to
either 
$ \theta_\infty+ \theta_0+\theta_1$ 
or to $2 d$, 
depending on the product of the signs of $\theta_1$ and $\theta_x$,
see Table \ref{TableConfluence},

The process could continue for $\PV \to \PIV$ or others,
but this does not happen.
Indeed,
this noncommutativity of the homographies and the confluence
never occurs at a lower level,
since it is already taken into account by the fact that $\PV$ has two
daughter $\Pn$ equations.

The two sequences of first degree {\BiT}s (normal and biased) are described
in the next two sections.
The monodromy exponents $\theta_j$ are defined in Table \ref{Table0}.
By convention, when the signs satisfy $s_j=S_j$,
the \BiT\ is equal to its inverse.

Let us now perform the standard confluence on $\TPVI$ and $\TPVIb$,
both for 
the affine        representation 
and
the birational    representation (\ref{eqTP6uvecTUnsigned}).
This will allow us to exhaust all the first degree {\BiT}s found by various 
authors.

On the affine matrix representation such as (\ref{eqT6biasedAffine0}),
the confluence acts as,
\begin{eqnarray}
& &
\matrix{
6 \to 5 :
& \hbox{lines }  (1,2,3+4) \to (1,2,3),\ \hfill
& \hbox{columns }(1,2,3)   \to (1,2,3),\
\hfill \cr
5 \to 4 :
& \hbox{lines }  (1+3,2) \to (1,2),\ \hfill
& \hbox{columns }(1,2)   \to (1,2),\
\hfill \cr
5 \to 3 :
& \hbox{lines }  (1+2,3) \to (1,2),\ \hfill
& \hbox{columns }(1,3)   \to (1,2),\
\hfill \cr
4 \to 2 :
& \hbox{lines }  (1+2) \to (1),\ \hfill
& \hbox{columns }(1)   \to (1),\ 
\hfill \cr
3 \to 2 :
& \hbox{lines }  (2) \to (1),\ \hfill
& \hbox{columns }(1) \to (1).
\hfill \cr
}
\end{eqnarray}

\subsection {The normal sequence}

This sequence is
\begin{eqnarray}
& &
\PVI\ : \
\pmatrix{s_\infty \theta_\infty \cr
         s_0      \theta_0      \cr
         s_1      \theta_1      \cr
         s_x      \theta_x      \cr}
= \frac{1}{2}               \pmatrix{ 1 & -1 & -1 & -1 \cr
                                     -1 &  1 & -1 & -1 \cr
                                     -1 & -1 &  1 & -1 \cr
                                     -1 & -1 & -1 &  1 \cr}
\pmatrix{S_\infty \Theta_\infty \cr
         S_0      \Theta_0      \cr
         S_1      \Theta_1      \cr
         S_x      \Theta_x      \cr}
              + \frac{1}{2} \pmatrix{ 1 \cr 1 \cr 1 \cr 1 \cr}\ccomma
\label{eqT6Affine}
\\
& &
\PV\ : \
\pmatrix{s_\infty \theta_\infty \cr
         s_0      \theta_0      \cr
         s_1      \theta_1      \cr}
= \frac{1}{2} \pmatrix {1 & -1 & -1 \cr -1 & 1 & -1 \cr -2 & -2 & 0 \cr}
\pmatrix{S_\infty \Theta_\infty \cr
         S_0      \Theta_0      \cr
         S_1      \Theta_1      \cr}
              + \frac{1}{2} \pmatrix{ 1 \cr 1 \cr 2 \cr}\ccomma\
s_\infty d = S_\infty D,
\label{eqT5Affine}
\\
& &
\PIV\ : \
\pmatrix{s_\infty \theta_\infty \cr
         s_0      \theta_0      \cr}
= \frac{1}{2}      \pmatrix{-1 & -3 \cr
                            -1 &  1 \cr}
\pmatrix{S_\infty \Theta_\infty \cr
         S_0      \Theta_0      \cr}
              + \frac{1}{4} \pmatrix{ 3 \cr 1 \cr}\ccomma\
s_\infty c = S_\infty C,
\label{eqT4Affine}
\\
& &
\PIII'\ : \
\pmatrix{s_\infty \theta_\infty \cr
         s_0      \theta_0      \cr}
=       \pmatrix{ 0 & -1 \cr -1 &  0 \cr}
\pmatrix{S_\infty \Theta_\infty \cr
         S_0      \Theta_0      \cr}
              +     \pmatrix{ 1 \cr 1 \cr}\ccomma\ 
s_\infty c = S_\infty C,\
s_0 d = S_0 D,
\\
& &
\PII\ : \
\pmatrix{s \theta_\infty \cr}= - \pmatrix{S \Theta_\infty \cr}
                               + \pmatrix{ 1 \cr}\ccomma\
s_\infty d = S_\infty D.
\label{eqT2Affine}
\end{eqnarray}
For the signs $s_j=1$,
all the shifts are positive,
and,
for the signs $s_j=S_j$,
the linear part has determinant $-1$.
The sum of the shifts remains equal to two
(except for $\PIV$ and $\PII$ because of a global rescaling,
see Table \ref{Table0}).

Since the affine representations (\ref{eqT6Affine})--(\ref{eqT2Affine})
are involutions when the signs satisfy $s_j =S_j$,
only half of the birational representation needs to be written,
the second half resulting from the permutation of 
$(u,\vect,c,d)$ and $(U,\vecT,C,D)$.

The confluence on $\TPVI$ results in 
(we omit the signs, they can easily be restored)
\begin{eqnarray}
\PVI\ : \
\frac{N}{u-U}
& = &
  \frac{x (x-1) U'}{U (U-1)(U-x)}
 +\frac{\Theta_0}{U}+\frac{\Theta_1}{U-1}+\frac{\Theta_x-1}{U-x}\ccomma
\label{eqTP6uvecT}
\\
\PV\ : \
\frac{N}{u-U}
& = &
  \frac{x U'}{U (U-1)^2}
 +\frac{\Theta_0}{U}+\frac{\Theta_1-1}{U-1}+\frac{D x}{(U-1)^2}\ccomma
\label{eqTP5uvecT}
\\
\PIV\ : \
\frac{N}{u-U}
& = &
  \frac{U'}{U}
 +\frac{4 \Theta_0}{U}+ C U + 2 C x,
\label{eqTP4uvecT}
\\
\PIII'\ : \
\frac{N}{u-U}
& = &
  \frac{xU'}{U^2}
 +\frac{\Theta_0-1}{U}+\frac{D x}{2 U^2}-\frac{C}{2}\ccomma
\label{eqTP3uvecT}
\\
\PII\ : \
\frac{N}{u-U}
& = &
  U' + D U^2 + D \frac{x}{2}\ccomma
\label{eqTP2uvecT}
\end{eqnarray}
with
\begin{eqnarray}
\PVI\ : \
N & = & 1-\Theta_\infty-\Theta_0-\Theta_1-\Theta_x
    = (1/2) \sum (\theta_j - \Theta_j),
\\
\PV\ : \
N & = & 1-\Theta_\infty-\Theta_0-\Theta_1
    = (1/2) \sum (\theta_j - \Theta_j),
\\
\PIV\ : \
N & = & -2(1-2 \Theta_\infty-2 \Theta_0)
    = 2 \sum (\theta_j - \Theta_j),
\\
\PIII'\ : \
N & = & 1-\Theta_\infty-\Theta_0
    = (1/2) \sum (\theta_j - \Theta_j),
\\
\PII\ : \
N & = & \frac{1}{2}-\Theta_\infty
    = (1/2) (\theta_\infty - \Theta_\infty).
\end{eqnarray}

These transformations were first found respectively,
for $\PV$ by Okamoto \cite{Okamoto1987II},
for $\PIV$  by Murata \cite{Murata1985},
for $\PIII$ by Fokas and Ablowitz \cite[Eq.~(4.4ab)]{FA1982},
for $\PII$  by Lukashevich \cite{Luka1971}.

\subsection {The biased sequence}

On the biased transformation $\TPVIb$ (also an involution),
the confluence yields
\begin{eqnarray}
& &
\PVI\ : \
\pmatrix{s_\infty \theta_\infty \cr
         s_0      \theta_0      \cr
         s_1      \theta_1      \cr
         s_x      \theta_x      \cr}
=-\frac{1}{2}              \pmatrix{ 1 & -1 & -1 & -1 \cr
                                    -1 &  1 & -1 & -1 \cr
                                    -1 & -1 &  1 & -1 \cr
                                    -1 & -1 & -1 &  1 \cr}
\pmatrix{S_\infty \Theta_\infty \cr
         S_0      \Theta_0      \cr
         S_1      \Theta_1      \cr
         S_x      \Theta_x      \cr}
              + \frac{1}{2} \pmatrix{-1 \cr 1 \cr 1 \cr -1 \cr}\ccomma
\label{eqT6biasedAffine}
\\
& &
\PV\ : \
\pmatrix{s_\infty \theta_\infty \cr
         s_0      \theta_0      \cr
         s_1      \theta_1      \cr}
=-\frac{1}{2} \pmatrix {1 & -1 & -1 \cr -1 & 1 & -1 \cr -2 & -2 & 0 \cr}
\pmatrix{S_\infty \Theta_\infty \cr
         S_0      \Theta_0      \cr
         S_1      \Theta_1      \cr}
+ \frac{1}{2} \pmatrix { -1 \cr 1 \cr 0 \cr},\
s_\infty d = - S_\infty D,
\\
& &
\PIV\ : \
\pmatrix{s_\infty \theta_\infty \cr
         s_0      \theta_0      \cr}
= - \frac{1}{2} \pmatrix { -1 & -3 \cr  -1 & 1 \cr}
\pmatrix{S_\infty \Theta_\infty \cr
         S_0      \Theta_0      \cr}
+ \frac{1}{4} \pmatrix {-1 \cr 1 \cr},\
s_\infty c = - S_\infty C,
\\
& &
\PIII'\ : \
\pmatrix{s_\infty \theta_\infty \cr
         s_0      \theta_0      \cr}
= -     \pmatrix{ 0 & -1 \cr
                 -1 &  0 \cr}
\pmatrix{S_\infty \Theta_\infty \cr
         S_0      \Theta_0      \cr}
\ccomma\ 
s_\infty c = - S_\infty C,\
s_0 d = - S_0 D,
\\
& &
\PII\ : \
\pmatrix{s \theta_\infty \cr}= \pmatrix{S \Theta_\infty \cr}\ccomma\
s_\infty d = - S_\infty D.
\label{eqT2biasedAffine}
\end{eqnarray}
At the $\PIII$ level,
the transformation reduces to the permutation of the two singular points
$(\infty,0)$,
a homography on $u$ which leaves $\PIII$ invariant.
Therefore, at the $\PII$ level this is just the identity.

The biased affine representations
(\ref{eqT6biasedAffine})--(\ref{eqT2biasedAffine})
and the normal ones 
(\ref{eqT6Affine})--(\ref{eqT2Affine})
have opposite linear parts 
(this results from our involution convention),
but the sum of the biased shifts is zero.

The biased birational transformations are
\begin{eqnarray}
\PVI
& : &
\frac{-N x (x-1)}{(u-x)(U-x)-x(x-1)}
= 
(U-x) \times
\nonumber
\\
& &
\phantom{xxxxxxxxxxxxxxx}
\left(  \frac{x (x-1) U'}{U (U-1)(U-x)}
 +\frac{\Theta_0}{U}+\frac{\Theta_1}{U-1}
 +\frac{\Theta_\infty-\Theta_0-\Theta_1}{U-x}
\right)
\ccomma
\label{eqbiraP6biased}
\\
\PV
& : &
\frac{-2 D x}{(u-1)(U-1)}
 =
(U-1) \left(
  \frac{x U'}{U (U-1)^2}
 +\frac{\Theta_0}{U}+\frac{\Theta_\infty-\Theta_0}{U-1}+\frac{D x}{(U-1)^2} 
\right)
\ccomma\
\\
\PIV
& : &
 2 C (u+U) = \frac{U'}{U} + \frac{4 \Theta_0}{U} + C U - 2 C x,
\label{eqTP4uvecTLuka}
\\
\PIII
& : &
\frac{D x}{u U}=-C,\
\\
\PII
& : &
u+U=0.
\end{eqnarray}
In (\ref{eqbiraP6biased}),
the constant $N$ is any expression among 
(\ref{eqP6Nbiased0})--(\ref{eqP6Nbiased4}).
The transformation for $\PV$ has first been obtained 
by Gromak \cite[Eq.~(13)]{Gromak1976},
and the one for $\PIV$ by Lukashevich \cite{Luka1967}.

Let us denote $\TPVIb$, $\TPVb$ and $\TPIVb$ the biased transformations,
              $\TPVIu$, $\TPVu$ and $\TPIVu$ the normal ones,
and $\H$ the unique homography of $\PV$ which conserves $x$,
\begin{eqnarray}
\PV
& : &
\H (x,u     ,\theta_\infty,\theta_0,\theta_1)
=  (x,u^{-1},\theta_0,\theta_\infty,\theta_1).
\end{eqnarray}

One has the relations
\begin{eqnarray}
& &
\TPVIu=\Sa \Sd \TPVIb \Hdcba \Sd \Sa,
\\
& &
\TPVu=\Sa \TPVb \Sa \Sc \TPVb \Sa \H,
\\
& &
\TPIVu=\Sa \TPIVb \Sb \TPIVb \Sa,
\label{eqT4ub}
\end{eqnarray}
(the relation (\ref{eqT4ub}) is due to \cite{BCH1995}).
At the $\PVI$ level, the relation is clearly invertible
but, at the $\PV$ and $\PIV$ levels,
we could not find inverse relations expressing the 
biased transformations as powers of the unbiased ones.
Therefore, the biased {\BiT}s are more elementary than the unbiased ones.


\section{Contiguity relation and its continuum limit}
\label{sectionDiscrete}

{}From each \BiT,
one easily deduces a contiguity relation,
which generalizes, as noted by Garnier \cite{Garnier1951},
that of the hypergeometric equation of Gauss.
Its systematic computation is as follows \cite{FGR}.

\begin{enumerate}
\item
Consider the \BiT, i.e.~the direct \BiT\ and its inverse
\begin{eqnarray}
& &
u=f(U,U',x,\vect,\vecT),\
\vect=g(\vecT),
\label{eqSTDirect}
\\
& &
U=F(u,u',x,\vecT,\vect),\
\vecT=G(\vect).
\label{eqSTInverse}
\end{eqnarray}

\item
Evaluate it at the values 
$(\vminus,v,\vplus)$ taken by a discrete variable 
at three contiguous points $(z-h,z,z+h)$, with $z=n h$,
\begin{eqnarray}
& &
 \vplus=f(v,v',x,f(\vect),\vect),\
\\
& &
\vminus=F(v,v',x,F(\vect),\vect).\
\end{eqnarray}

\item
Eliminate the variable $v'$ between these two relations,
\begin{eqnarray}
& &
G(\vplus,\vminus,v,x,\vect)=0.
\end{eqnarray}

\end{enumerate}

For $\PVI$,
equations (\ref{eqSTDirect})--(\ref{eqSTInverse}) are equivalent to
\begin{eqnarray}
& &
\frac{x(x-1) U'}{U(U-1)(U-x)} =
 2 \frac{s_j \theta_j - S_j \Theta_j}{u-U}
-\left(
\frac{S_0\Theta_0}{U}+\frac{S_1\Theta_1}{U-1}+\frac{S_x\Theta_x-1}{U-x}\right),
\\
& &
\frac{x(x-1) u'}{u(u-1)(u-x)} =
-2 \frac{S_j \Theta_j - s_j \theta_j}{u-U}
-\left(
\frac{s_0\theta_0}{u}+\frac{s_1\theta_1}{u-1}+\frac{s_x\theta_x-1}{u-x}\right),
\end{eqnarray}
in which $j$ is anyone of the four singular points $(\infty,0,1,x)$,
and the contiguity relation is
\begin{eqnarray}
& &
\frac{\varphi(n+1/2)}{\vplus - v} + \frac{\varphi(n-1/2)}{\vminus - v}
=
 \frac{s_0 \theta_0 - S_0 \Theta_0}{v}
+\frac{s_1 \theta_1 - S_1 \Theta_1}{v-1}
+\frac{s_x \theta_x - S_x \Theta_x}{v-x}\ccomma
\label{eqContiguity}
\\
& &
\varphi(n)= 
        \frac{1}{2}
(s_\infty \theta_\infty + s_0 \theta_0 + s_1 \theta_1 + s_x \theta_x -1),
\end{eqnarray}
in which $\vect$ is taken at the center point $z=z_0 + n h$.
This very simple expression is clearly invariant under any permutation of 
the four singular points of $\PVI$.

This contiguity relation (\ref{eqContiguity}) can be interpreted as a 
second order discrete equation \cite{FGR}.
The two-point recurrence relation (\ref{eqT6}) 
admits five classes of solutions.
Each class, characterized by a signature,
leads to a different contiguity relation (\ref{eqContiguity}),
i.e.~to a different second order discrete equation.
Four of them are autonomous 
(signatures $(s_j S_j)=$ $(---+)$, $(--++)$, $(-+++)$, $(++++)$),
they cannot admit a continuum limit to a Painlev\'e equation.
The only nonautonomous one (signature $(----)$)
is
\begin{eqnarray}
& &
\frac{n+1/2}{\vplus - v} + \frac{n-1/2}{\vminus - v}
=
 \frac{n + K_2 (-1)^n}{v}
+\frac{n + K_3 (-1)^n}{v-1}
+\frac{n + K_4 (-1)^n}{v-x}\ccomma
\label{eqContigP6a}
\\
& &
K_2=- k_2 + k_3 + k_4,\
K_3=  k_2 - k_3 + k_4,\
K_4=  k_2 + k_3 - k_4.
\label{eqContigP6b}
\end{eqnarray}
In the continuum limit,
among the six simple poles of $v$ in the sum (including $\infty$),
the first two will create a second order derivative 
and the four others will define at most four singular points.
Since none of the last four poles depends on $n$,
it is impossible that the continuum limit be $\PVI$.

The transform of this discrete equation under
\begin{eqnarray}
(\vplus,v,\vminus) \mapsto (\vplus,x/v,\vminus)
\end{eqnarray}
has already been obtained \cite[Eq.~(1.5)]{NRGO}    
as a reduction of a lattice KdV equation,
together with a discrete Lax pair 
and a continuum limit to the full $\PV$. 
This is in agreement with the continuum limit of the hypergeometric
contiguity relation,
which is not the hypergeometric equation but a confluent one.
Nevertheless,
we do not know of a general proof of this feature.

\section{Conclusion}

The fundamental homography between the derivative of the solution of the
Painlev\'e equation and the Riccati pseudopotential
has allowed us to define a consistent truncation.
For $\PVI$, this truncation provides one first degree {\BiT}.
Under the confluence to the lower $\Pn$ equations,
this transformation generates 
two inequivalent first degree {\BiT}s at the $\PV$ and $\PIV$ level,
and only one at the $\PIII$ and $\PII$ level.
This provides a coherent description of all the first degree {\BiT}s of all 
the $\Pn$ equations which admit one.

As to the contiguity relation of the first degree \BiT\ of $\PVI$,
it defines exactly one nonautonomous second order difference equation.
If again one applies to this single discrete equation
the two noncommuting operations of confluence and homography
which have generated the normal and biased sequences of {\BiT}s,
one should similarly define a coherent description including many of the
currently existing discrete Painlev\'e equations.


\section*{Acknowledgments}

The authors warmly thank J.~Satsuma and T.~Tokihiro
for their generous hospitality,
and they acknowledge the financial support of the Tournesol grant T99/040.
RC thanks the French ``minist\`ere des affaires \'etrang\`eres'' 
for travel support.



\vfill \eject
\end{document}